\documentclass[oneside,onecolumn,amsmath]{revtex4}
\usepackage{graphicx}
\usepackage{bm}
\usepackage{epsfig}
\usepackage{amssymb}
\usepackage{amsfonts}
\usepackage{braket}
\usepackage{color}

\usepackage{hyperref}
\usepackage{float}
\usepackage{bibentry}
\usepackage{epstopdf}

\begin{document}
\title{Chirality-driven Hall Transport Phenomena of Spins}
\author{Jung Hoon Han}
\email[Electronic address:$~~$]{hanjh@skku.edu}
\author{Hyunyong Lee}
\email[Electronic address:$~~$]{hyunyong.rhee@gmail.com}
\affiliation{Department of Physics, Sungkyunkwan University, Suwon 16419, Korea}
\date{\today}

\begin{abstract}
Experimental and theoretical aspects of Hall-type transport of spins in magnetic insulators are reviewed, with emphasis on their spin chirality origin. A general formalism for linear response theory of thermal Hall transport in the spin model is developed, which can be applied to both the magnon and the paramagnetic, spin-liquid regimes. Recent experiments on magnon-mediated thermal Hall transport in the two-dimensional kagome, and three-dimensional pyrochlore ferromagnetic insulators are reviewed in light of the multi-band magnon theory of Hall transport, and compared to the more mysterious thermal Hall transport found in the putative quantum spin ice material. As realizations of spin-chirality driven magnon transport in the real space, we review the general theory of emergent gauge fields governing the magnon dynamics in the textured magnet, and discuss its application to the magnon-Skyrmion scattering problem. Topological magnon Hall effect driven by the Skyrmion texture is discussed.
\end{abstract}

\maketitle

\section{Introduction}

A powerful concept that threads the spin liquid phase in both its classical and quantum forms is that of spin chirality, specifically the scalar spin chirality. Originally proposed as a tool to characterize a novel time-reversal symmetry-broken ground state of high-$T_c$ cuprates~\cite{laughlin,wen}, it is formed from the configuration of three neighboring spins $\vec{S}_i, \vec{S}_j, \vec{S}_k$ at the $\langle ijk\rangle$ sites by their triple product $\chi_{ijk} = \vec{S}_i \cdot (\vec{S}_j \times \vec{S}_k )$. It is odd, by construction, under the time-reversal operation that takes the spin $\vec{S}$ to its negative, $-\vec{S}$, and has zero average value for states that respect the time-reversal symmetry.

For the magnetically disordered state  where $\langle \vec{S}_i \rangle = 0$, one might conclude that spin chirality must also be zero, $\langle \vec{S}_i \cdot (\vec{S}_j \times \vec{S}_k ) \rangle = \langle \vec{S}_i \rangle \cdot \langle \vec{S}_j \rangle \times \langle \vec{S}_k \rangle = 0$, by virtue of the mean-field argument. This is not always the case, however, and one good way to identify a spin liquid is as a spin-disordered phase in which the chirality remains ordered, or at least have a much more extended correlation length than is expected in the uncorrelated, paramagnetic state. 

In this review we will take look at the spin chirality from a pragmatic vantage point, focusing on experimentally accessible consequences of spin states with significant amount of spin chirality order or correlations. Specifically we are going to focus on insulating magnets where conduction electrons as carriers of energy and spin are absent and the sole transport carriers are the spins. A number of recent experiments have found thermal Hall conductivity in insulating magnets with ferromagnetic or antiferromagnetic couplings, in both ordered and disordered regimes of the phase space~\cite{tokura10,tokura12,ylee15a,ylee15b,ylee16,ong15,matsuda-PNAS}. Some of these materials have the ferromagnetic ground state~\cite{tokura10,tokura12,ylee15a,ylee15b,ylee16}, others are believed to be in the quantum spin ice phase~\cite{ong15} or in the spin liquid phase~\cite{matsuda-PNAS}. We will outline a linear-response formulation of the thermal transport suggesting that Hall effects in purely spin systems must be closely tied to significant spin chirality correlations. Often, the symmetry is broken and the spin chirality is reduced to a simpler order. For one, a non-coplanar spin order can give $\chi_{ijk} = \langle \vec{S}_i \rangle \cdot \langle \vec{S}_j \rangle \times \langle \vec{S}_k \rangle$ purely from mean-field argument. Under an applied magnetic field $\vec{B}$, scalar spin chirality turns into a vector spin chirality $\chi_{ijk} \rightarrow \vec{B} \cdot (\vec{S}_j \times \vec{S}_k ) + ({\rm cyc. ~ perm.})$. Despite the differences, we will allude to Hall-like transport in all these systems as chirality-driven. 

Recent rapid advances on Skyrmion research~\cite{pfleiderer,tokura-skyrmion,nagaosa-review,recent-Skyrmion} has spurred interests in magnon transport through the Skyrmionic spin texture background~\cite{liu11,jiadong13,loss13,batista14,kovalev14,iwasaki14,garst14,mochizuki14,yt15}. Since Skyrmion is but a quantized form of spin chirality which acts as a source of emergent magnetic flux on the magnon particles, magnon transport in the Skyrmion matter is an excellent example of chirality-driven spin transport phenomena which is gaining attention as a new frontier in magnonics~\cite{magnonics-review,bauer,maekawa}. 

The transport manifestation of the spin chirality can be pictured as follows. Say a given cluster of spins fluctuate in such manner as to maintain a definite spin chirality. If one of the spins in that cluster is perturbed, the other two spins must be moved in some coordinated dance, such that the same spin chirality is maintained at all times. Viewed from the apex of the three-spin umbrella, such a collective motion should appear as either clockwise or counter-clockwise rotation of the spin cluster around the umbrella's edge. There is a sense of vorticity built into such a constrained motion, which translates ultimately into a magnetic flux of emergent nature for the spin excitation.  This kind of emergent magnetism can be manifested as the formation of topologically non-trivial bands of magnons~\cite{KNL,murakami11,murakami14,TMI,mook,lee15,kovalev16,tserkovnyak16,owerre} in the momentum space, or as localized emergent magnetic fields in real space~\cite{liu11,jiadong13,loss13,batista14,kovalev14,iwasaki14,garst14,mochizuki14,yt15}. While ideas of this sort have been around for some time now, what makes these concepts exciting now and warrants the writing of a review is the interesting breakthroughs that have taken place over the past few years, which have made them real and experimentally accessible concepts. 

We formulate the rest of the review by starting with a theory of thermally driven spin transport in purely spin systems and cascading down to the magnon picture by making mean-field approximations. Contacts with many existing magnon-based theories of thermal Hall transport can be established this way. Then we will cover several exciting experiments that came out in the last few years in light of the theoretical picture. In the last phase of the review we discuss the real-space manifestation of the chirality-driven magnon dynamics in the context of Skyrmion matter. A summary and perspective section is devoted to the very latest development on the experimental front and some speculations on the future of theory. 

\section{Linear Response Theory of Spin and Energy Currents in Spin Systems}

\subsection{Formal matters}
We are concerned with developing a theory of spin transport in purely spin systems, mostly focused on, but by no means confined to magnetically ordered phases where magnons are the quasi-particle carriers of spin, energy, and possibly information itself. Electrical currents in a conductor are generated by a voltage difference, irradiation by electromagnetic wave, and temperature or chemical potential gradients. Spin excitations in insulating magnets are made possible by coupling to the electromagnetic wave, temperature gradient, and possibly the magnetization gradient. The generation of spin and heat currents has been successfully demonstrated by microwave~\cite{saitoh10a} and temperature gradient\cite{saitoh10b} techniques, calling for developments of theory of transport in magnetic insulators~\cite{maekawa}. In this review we will be particularly concerned with spin transport in {\it topological magnets}, in which some form of topological effects influence the spin dynamics. 

There are several ways of developing a transport theory. One can start from the microscopic Landau-Lifshitz-Gilbert dynamical equations and work one's way up to derive some phenomenological equations of magnons, as exemplified in the work of Ref. \onlinecite{tserkovnyak12}. Alternatively, there is a linear response approach set forth by Kubo, in which the aim is to obtain transport coefficients directly from the underlying microscopic model. Transport coefficients such as the thermal Hall conductivity are written as integrals of some equilibrium-averaged physical quantities. We will discuss how the linear response approach can be applied to derive transport coefficients when the only physical degrees of freedom are spins.

In the linear response theory one (i) begins with a proper definition of the current operator for a given microscopic model, and (ii) identifies the classical source that couples to it. For the electronic system this is done by asking how the density operator at the site $i$ (omitting spin indices) $\rho_{i} = c^\dag_{i}c_{i}$ evolves over time

\begin{equation} \dot{ \rho}_{i} = { i \over \hbar } [H, \rho_{i} ]  = - \sum_{j \neq i} J_{i\rightarrow j} . \end{equation}
In all subsequent formulas we set $\hbar =1$. Evaluating the commutator on the right hand side amounts to working out the lattice divergence of the current operator $J_{i\rightarrow j}$, defined as the flow from site $i$ to $j$. Thus, by a straightforward application of the commutator algebra one obtains the current operator $J_{i \rightarrow j}$, which completes the task (i) of the linear response program.  

The spin current operator projected along, say, the $z$-direction can be deduced in the same way. We define as the spin density the local operator $S^z_i$, and ask to derive the Heisenberg equation of motion for it:

\begin{equation} \dot{S}_i^z = {i } [H, S_i^z ] = -  \sum_{j} J^S_{i \rightarrow j} . \end{equation}
For the Heisenberg spin exchange model $H= J \sum_{\langle ij\rangle} \vec{S}_i \cdot \vec{S}_j$ one finds

\begin{equation} J^S_{i\rightarrow j} = J \hat{z} \cdot (\vec{S}_i \times \vec{S}_j )  = - J^S_{j\rightarrow i} .  \end{equation}
This simple result, applicable to any lattice geometry, says that the spin current operator of the Heisenberg spin model is a sum of vector spin chiralities. 

The source term that couples to the spin density is the Zeeman field. Adding the Zeeman term  $\sum_i B_i S_i^z$ to the spin Hamiltonian modifies the initial density matrix $\rho_0$ by an amount $\delta\rho$, given by~\cite{kubo57,luttinger64} 

\begin{equation} \delta\rho = -\rho_0 \int_0^\infty dt' e^{-st'} \int_0^{\beta} d\beta' \sum_{\langle ij \rangle} (B_j - B_i ) J^S_{i\rightarrow j} (-t' - i\beta'),  \end{equation}
where $s$ is an arbitrary small number. The equilibrium average of the current operator is zero, but the one with respect to $\delta\rho$ is not, and gives

\begin{eqnarray} && \langle J^S_{i \rightarrow j} \rangle = - \int_0^\infty dt' e^{-st'} \int_0^{\beta} d\beta' \nonumber \\
&& ~~~~~~~~~~ \times \sum_{\langle kl \rangle } {\rm Tr} \left[ \rho_0 J^S_{i\rightarrow j} J^S_{k\rightarrow l} (-t' - i\beta' ) \right] (B_l - B_k ) .  \end{eqnarray}
Summing over all bond current operators in a particular orientation $\sum_i \langle J^S_{i\rightarrow i + e_a} \rangle = \langle  I^S_a \rangle $, for a given magnetic field gradient in the $b$-direction $l=k+e_b$, $B_l - B_k \simeq \partial_b B$, yields the linear-response formula

\begin{eqnarray} & I^S_a = \kappa^S_{ab} (- \partial_b B ) ,  \nonumber \\
& \kappa^S_{ab} = \int_0^\infty dt' e^{-st'} \int_0^\infty  d\beta' {\rm Tr} \left[ \rho_0 I^S_{a} I^S_{b} (-t' - i\beta' ) \right] .  \label{eq:spin-Hall} \end{eqnarray}
The coefficients $\kappa^S_{ab}$ are a measure of the spin current flowing in response to the magnetic field gradient. It is closely related to the vector spin chirality correlation function. 

Several key experiments in the last few years have revealed that a transverse current of heat is generated when a temperature gradient is imposed across certain magnetic insulators (meaning insulators with only magnetic degrees of freedom, but not necessarily magnetically ordered)~\cite{tokura10,tokura12,ylee15a,ylee15b,ylee16,ong15,matsuda-PNAS}. Heat (energy) flow in response to, and along the direction of, the temperature gradient is expected from the second law of thermodynamics. On the other hand, spin-mediated flow of heat transverse to the temperature gradient requires a physical mechanism akin to the Lorentz force on electrons. We will feed this question to the linear response program. 

When the Hamiltonian consists of a collection of local terms, $H = \sum_i H_i$, and each $H_i$ consists of the spin $\vec{S}_i$ and a finite number of other spins $\vec{S}_j$ in its neighborhood, it is reasonable to choose $H_i$ as the local energy density operator. The right definition of the energy current operator emerges from the commutator:

\begin{equation} \dot{H}_i = {i} [H , H_i ] = {i} \sum_{j\neq i} [H_j , H_i ] = - \sum_{j \neq i } J^E_{i\rightarrow j} . \end{equation}
The commutator $[H_j, H_i]$ vanishes unless some spin operator belongs to both $H_i$ and $H_j$. For the Heisenberg spin model, the local Hamiltonian is $H_i = (J/2) \sum_{j\in i} \vec{S}_i \cdot \vec{S}_j$, where $j\in i$ denotes spins $\vec{S}_j$ in the immediate neighborhood (nearest neighbors) of the spin $\vec{S}_i$. Working out the commutator in question gives
out the energy current operator 

\begin{equation} J^E_{i \rightarrow j} = {1\over 2} J^2  \sum_{k\in j} \chi_{ijk} + {1\over 2} J^2 \sum_{k \in i} \chi_{ijk}  = -J^E_{j\rightarrow i} . \label{eq:J-E} \end{equation}
It is given as a collection of scalar spin chiralities over the neighboring sites of the $\langle ij\rangle$ bond. 

The energy density $H_i$ couples to a ``gravitational" potential source $\psi_i$ and adds a term $\sum_i \psi_i H_i$ to the spin Hamiltonian~\cite{luttinger64,obraztsov,streda77,murakami11,murakami14}. The density matrix operator picks up a correction,

\begin{equation}\delta\rho \simeq -\rho_0 \int_0^\infty dt' e^{-st'} \int_0^\beta d\beta' \sum_{\langle ij \rangle} (\psi_j - \psi_i ) J^E_{i\rightarrow j} (-t' - i\beta') .  \end{equation} 
Average of the energy current operator $I^E_a = \sum_{i} \langle J^E_{i\rightarrow i+e_a} \rangle$ in response to the gradient $\partial_b \psi$ is

\begin{eqnarray} & I^E_a = \kappa^E_{ab} (- T \partial_b \psi ) ,  \nonumber \\
& \kappa^E_{ab} = {1\over T} \int_0^\infty dt' e^{-st'} \int_0^\beta  d\beta' {\rm Tr} \left[ \rho_0 I^E_{a} I^E_{b} (-t' - i\beta' ) \right] . \end{eqnarray}
Multiplication and division by the temperature factor $T$ is needed to convert the gravitational potential gradient to the temperature gradient~\cite{luttinger64}. The resulting quantity $\kappa^E_{ab}$ is the thermal conductivity tensor for spin models. 

It turns out, however, this is not the full answer to the energy response function. The source term we added modifies the local energy density operator to $(1+\psi_i ) H_i$. Running through the commutator algebra once again, we find the current operator in the presence of the source term is modified as well,

\begin{eqnarray} J^E_{i\rightarrow j} [\psi]  \simeq (1+\psi_i + \psi_j ) J^E_{i\rightarrow j} , \end{eqnarray}
neglecting second-order terms in $\psi$. While ${\rm Tr}[\rho_0 J^E_{i\rightarrow j} ] = 0$ as a condition for equilibrium, the correction term to the energy current does not necessarily vanish even in equilibrium:

\begin{equation} {\rm Tr} [\rho_0 ( \psi_i + \psi_j ) J^E_{i \rightarrow j } ]  = (\vec{\nabla} \psi )  \cdot {\rm Tr} 
[\rho_0 (\vec{r}_i + \vec{r}_j ) J^E_{i \rightarrow j} ]  \neq 0 . \end{equation}
The operator inside the trace is reminiscent of the orbital magnetic moment $\vec{m} = \vec{r} \times \vec{J}$ and for this reason its average is known as the orbital magnetization current in the context of electronic thermal transport~\cite{luttinger64,obraztsov,streda77}, and rotating magnon current for magnon systems~\cite{murakami11}. The true thermal transport coefficient  is a sum of two contributions, one from the modification of the density matrix and the other from the modified definition of the energy current operator~\cite{lee15}:

\begin{eqnarray} 
	&&\kappa^E_{ab} = {1\over T} \int_0^\infty dt' e^{-st'} \int_0^\beta  d\beta' {\rm Tr} \left[ \rho_0 I^E_{a} I^E_{b} (-t' - i\beta' ) \right]\nonumber\\ 
	&&~~~~~~~~
	+ \frac{1}{T} {\rm Tr}\left( \rho_0 \left[ \frac{\partial I_a^E (\vec{q})}{\partial q_b}\right]_{\vec{q}=0} \right) .
\end{eqnarray}
Here $I_a^E(\vec{q})$ is the energy current operator in the Fourier space with momentum $\vec{q}$. In contrast to the thermal conductivity, the source term $\sum_i B_i S^z_i$  does not modify the spin current operator since the added term to the Hamiltonian commutes with $S_i^z$. The spin response formula $\kappa^S_{ab}$ remains as given in Eq. (\ref{eq:spin-Hall}).

Another quantity of much physical importance is the amount of spin flow $J^z_a$ induced by the thermal gradient $-T \partial_b \psi$. Running through the calculation that uses the gravitational source and the spin current response yields the spin Nernst coefficient~\cite{bauer}

\begin{eqnarray}
	&& \kappa^N_{ab} = {1\over T} \int_0^\infty dt' e^{-st'} \int_0^\beta  d\beta' {\rm Tr} \left[ \rho_0 I^S_{a} I^E_{b} (-t' - i\beta' ) \right]\nonumber\\ 
	&&~~~~~~~~
	+ \frac{1}{T} {\rm Tr}\left( \rho_0 \left[ \frac{\partial I_a^S (\vec{q})}{\partial q_b}\right]_{\vec{q}=0} \right).  \label{eq:full-kappa-E}
\end{eqnarray}
All three spin response functions $\kappa^S_{ab}, \kappa^E_{ab}, \kappa^N_{ab}$ derived here apply equally well to disordered spin systems as well as in their ordered phases. 

Progress in our understanding of chirality-driven Hall transport in spin systems, spurred in particular by the work of Ref.~\onlinecite{KNL}, has taught us that the Dzyaloshinskii-Moriya (DM) interaction plays an analogous role to magnons as the Aharonov-Bohm flux for electrons. 
A minimal model for investigating the thermal Hall effect of magnons has been developed over the years~
\cite{KNL,TMI,mook,lee15,kovalev16}, which consists of Heisenberg-Dzyaloshinskii-Moriya-Zeeman (HDMZ) terms, $H=\sum_i H_i$, 

\begin{equation} H_i = - {1\over 2}J  \sum_{j\in i} \vec{S}_i \cdot \vec{S}_j + {1\over2} \sum_{j\in i} \vec{D}_{i\rightarrow j} \cdot \vec{S}_i \times \vec{S}_j - \vec{B} \cdot \vec{S}_i . \label{eq:hdmz} \end{equation}
There are three sets of parameters $J, \vec{D}_{i\rightarrow j} = - \vec{D}_{j\rightarrow i}$, $\vec{B} = g\mu_B \vec{H}$ having the dimension of energy, and the dimensionless spin operator  normalized to $\vec{S}\cdot \vec{S} = S(S+1)$. The spin and energy current operators for this model have been worked out in Ref. \onlinecite{lee15},

\begin{eqnarray} 
	&& J^z_{i\rightarrow j} = -i {J' \over 2} e^{i\phi_{i\rightarrow j}} S^\dag_i S^-_j + h.c. = -J^z_{j\rightarrow i} , \nonumber \\ 
	&& J^E_{i\rightarrow j} = - {1\over 2} \sum_{k \in j} \Bigl(J S^z_k J^z_{i \rightarrow j} + J S^z_i J^z_{j\rightarrow k} + [ J^z_{i\rightarrow j}, J^z_{j\rightarrow k} ] \Bigr) \nonumber \\
	&& ~~~ + {1\over 2} \sum_{k \in i} \Bigl(J S^z_k J^z_{j \rightarrow i} + J S^z_j J^z_{i\rightarrow k} + 
[ J^z_{j\rightarrow i}, J^z_{i\rightarrow k} ] \Bigr) -B J^z_{i\rightarrow j} \nonumber \\
	&& ~~~~~~ = - J^E_{j\rightarrow i} , \label{eq:17}
\end{eqnarray} 
assuming all DM vectors parallel to the  $z$-axis, $\vec{D}_{i\rightarrow j} = D_{i\rightarrow j} \hat{z}$ and all $D_{i\rightarrow j}$'s having the same magnitude. The phase angle $\phi_{i\rightarrow j}$ comes from $\tan (\phi_{i\rightarrow j}) = D_{i\rightarrow j}/J$, and  $J' =\sqrt{J^2 + D^2}$ is the effective exchange energy. 

The energy current for the HDMZ Hamiltonian is extremely lengthy. Even for the much simpler energy current operator of the pure Heisenberg model, working out the correlation functions is a non-trivial task. In the following subsection we show how the calculation of various response functions is made tractable in the mean-field theory. 

\subsection{Mean-field theory}

A mean-field picture of both spin and energy current operators follows from Holstein-Primakoff replacement of spin operators $S^+_i = \sqrt{2S} b_i$, $S^-_i = \sqrt{2S} b^\dag_i$, assuming uniform average spin polarization $\langle \vec{S}_i \rangle = S\hat{z}$, in Eq. (\ref{eq:17}):

\begin{eqnarray} J^z_{i \rightarrow j } &\rightarrow& i J' S^{2} ( e^{i\phi_{i\rightarrow j}} b^\dag_j b_i - e^{i\phi_{j\rightarrow i}}  b^\dag_i b_j  )  , \nonumber \\
\chi_{ijk} & \rightarrow & i S^2 (b^\dag_j b_i - b^\dag_i b_j ) + (\mathrm{ cyc. ~ perm.}).  \end{eqnarray}
It is noteworthy that even with the DM interaction present, the ground state of the HDMZ Hamiltonian is still that of a perfect ferromagnet without canting provided the local DM vectors satisfy the sum rule: $\sum_{j\in i} \vec{D}_{i\rightarrow j} = 0$.

The spin current becomes the boson current operator. The energy current is quite complicated to write down, even at the mean-field level.  The spin Hamiltonian is replaced by its Holstein-Primakoff approximation~\cite{lee15},

\begin{equation} H \rightarrow H_{\rm MF} = \sum_{\vec{ k} } \Psi^\dag_{\vec{k}} H_{\vec{k}} \Psi_{\vec{k}}, \label{eq:H-MF}\end{equation}
using a set of Holstein-Primakoff boson operators organized as an $N$-dimensional column vector $\Psi_{\vec{k}}$ and a $N\times N$ matrix $H_{\vec{k}}$. The magnon Hamiltonian used by many authors~\cite{murakami11,TMI,mook,lee15,kovalev16} can be recovered this way. Schwinger boson theory does a similar job of reducing the Hamiltonian to the mean-field form~\cite{lee15,tserkovnyak16,owerre}, with the advantage over the Holstein-Primakoff scheme that no preferred magnetization direction needs to be assumed. In both schemes, the spin and energy current operators become~\cite{lee15}

\begin{eqnarray} \vec{j}^S &=& \sum_{\vec{k}} \Psi^\dag_{\vec{k}} {\partial H_{\vec{k}} \over \partial \vec{k} } \Psi_{\vec{k}} ,\nonumber \\
\vec{j}^E &=&{1\over 2} \sum_{\vec{k}} \Psi^\dag_{\vec{k}} \left( {\partial H_{\vec{k}} \over \partial \vec{k} } H_{\vec{k}} +  H_{\vec{k}}  {\partial H_{\vec{k}} \over \partial \vec{k} } \right) \Psi_{\vec{k}}   . \label{eq:j-MF} \end{eqnarray}

One of the attractive features of the mean-field approach, apart from the ease of evaluating correlation functions, is that it facilitates the development of the theory of spin dynamics in analogy with those of electrons on the same lattice. For instance the mean-field Hamiltonian (\ref{eq:H-MF}) and the current operators in Eq. (\ref{eq:j-MF}) can equally be those of the electronic model. Diagonalizing the matrix $H_{\vec{k}}$ gives $N$ bands and the Berry flux distribution function $\Omega_{\vec{k}}^{(n)}$ for each $n$-th band, $1\le n \le N$. One important modification in the boson case arises due to the occupation number function being Bose-Eisntein rather than Fermi-Dirac. Boson numbers are not conserved and vanish at low temperature, which also makes a big difference in the sensitivity of the response functions on temperature. Zeeman energy can easily tune the energy gap in the boson spectrum and affect the thermal response substantially, when the mean-field bandwidth of bosons set by the exchange energy $J$ is weak and comparable to the laboratory field strengths reaching 10T. All in all, when thinking about the response functions in spin systems it is useful to keep in mind that features of the boson bands are more ``fragile" than those of fermions. 

\begin{figure}
\begin{center}
\includegraphics[width=0.4\textwidth]{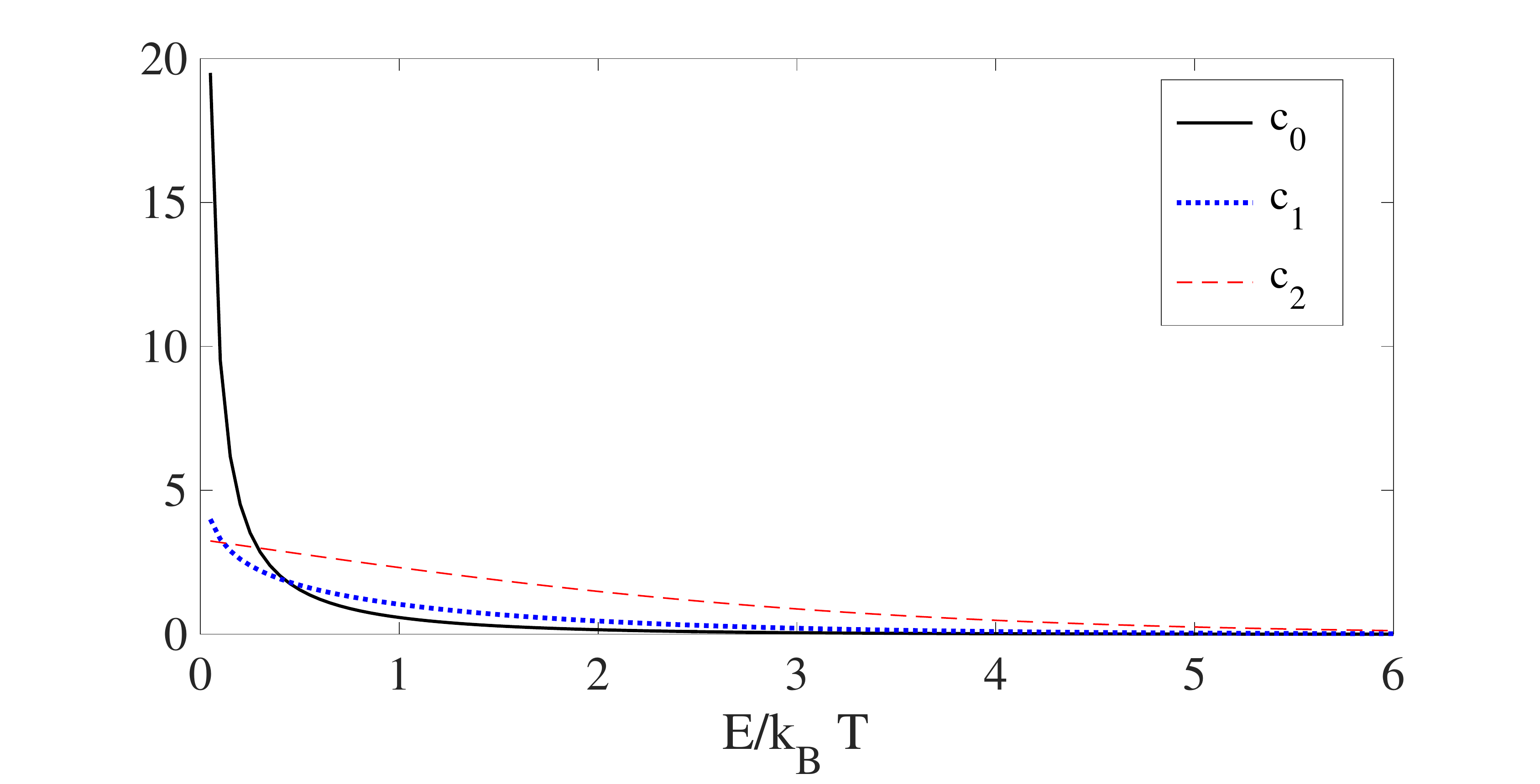}
\caption{ Comparison of the $c_0$\,(Bose-Einsten function), $c_1$ and $c_2$ functions for the same reduced energy $x=E/k_B T$.}
\label{fig:1}
\end{center}
\end{figure}	

In the mean-field picture the three kinds of Hall-type response functions mentioned in the previous section become~\cite{murakami11,lee15,kovalev16}

\begin{eqnarray} &&  \kappa_{xy}^S = {\mu_B \over \hbar V} \sum_{\vec{k}, n}  c_0 \bigl(B^{(n)}_{\vec{k}} \bigr)  \Omega^{(n)}_{\vec{k}} ,  \nonumber \\
& & \kappa_{xy}^N = {k_B T \over \hbar V} \sum_{\vec{k}, n}  c_1 \bigl(B^{(n)}_{\vec{k}} \bigr)  \Omega^{(n)}_{\vec{k}} ,  \nonumber \\
& & \kappa_{xy}^E = - {k_B^2 T \over \hbar V} \sum_{\vec{k}, n} c_2 \bigl(B^{(n)}_{\vec{k} } \bigr) ~  \Omega^{(n)}_{\vec{k}} ,
\label{eq:kappa-MF} \end{eqnarray}
as the sum over all $1\le n \le N$ bands and over momenta $\vec{k}$ of the first Brillouin zone, divided
by the number of unit cells $N$ in the lattice.  The three weight functions $c_0, c_1, c_2$ are given by~\cite{murakami11}

\begin{eqnarray} 
& c_0 (B) = B , \nonumber \\
& c_1 (B) = (1+B) \ln (1+B) - B \ln B , \nonumber \\
& c_2 ( B ) = (1+B) \left( \ln {1+B \over B} \right)^2 - (\ln B)^2 -2 \rm{Li}
(-B),
\end{eqnarray}
in terms of the Bose-Einstein function $B (E) = \left( e^{E/k_B T} - 1 \right)^{-1}$ and plotted in Fig. \ref{fig:1} for comparison. 
One sees that higher-numbered weight functions tend to be broader and probe the higher energy region more. 
The two response functions $\kappa^S_{xy}$ and $\kappa^E_{xy}$ were calculated for the kagome lattice in Ref. \onlinecite{lee15}. The spin Nernst coefficient $\kappa^N_{xy}$ for the kagome~\cite{kovalev16} and honeycomb~\cite{tserkovnyak16,owerre} lattices have been calculated recently. 

\begin{figure}
\begin{center}
\includegraphics[width=0.4\textwidth]{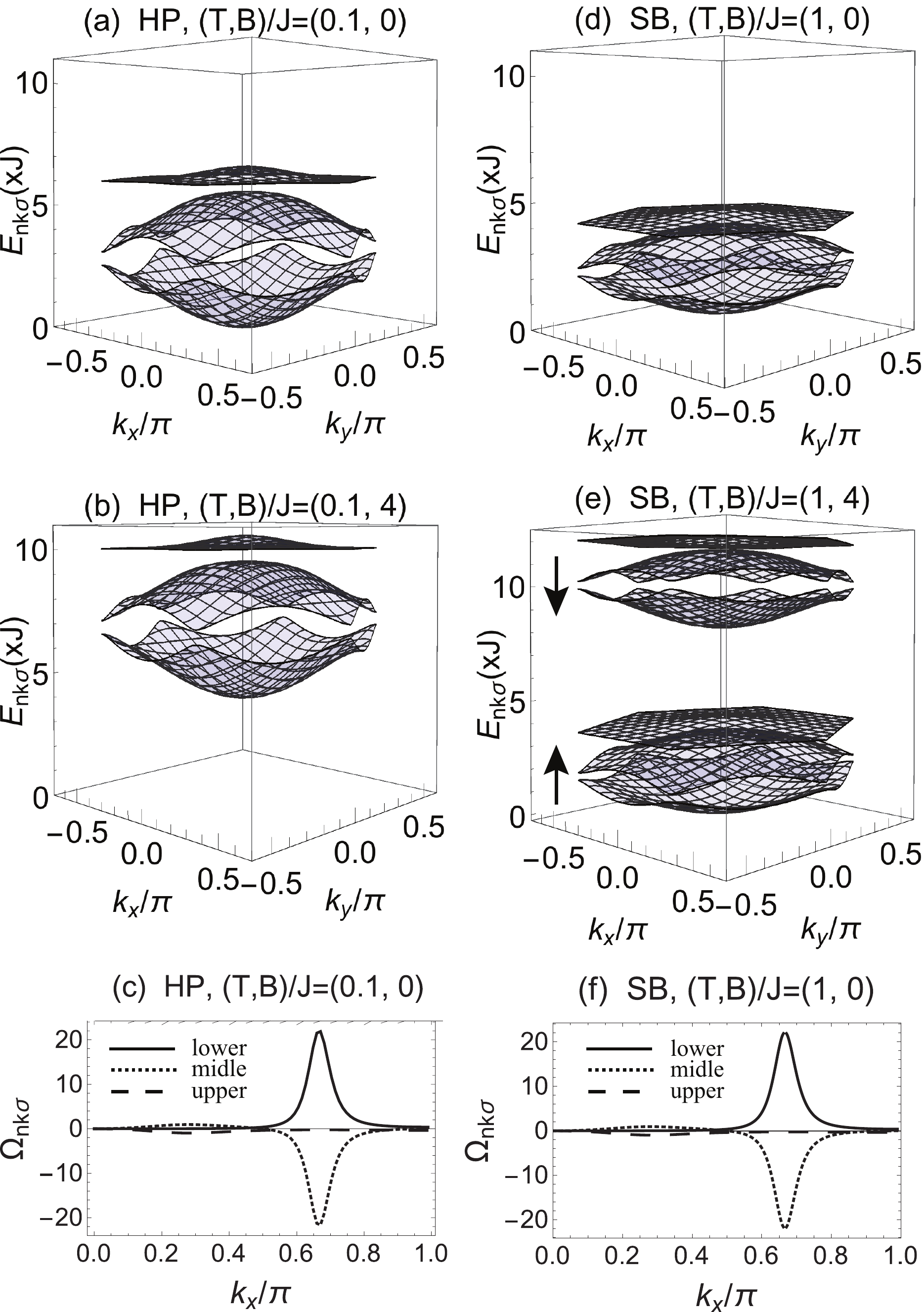}
\caption{ Mean-field band structures from Holstein-Primakoff [(a),(b)] and Schwinger boson [(d),(e)] reduction of the HDMZ spin Hamiltonian. Berry flux plotted along the $k_y = 0$ line is shown for HP [(c)] and SB [(f)] models (reproduced with permission from Ref. \onlinecite{lee15}.)}
\label{fig:2}
\end{center}
\end{figure}

The magnon band structure and the accompanying Berry flux distribution of the HDMZ Hamiltonian were worked out in Refs. \onlinecite{TMI,mook,lee15}. As shown in Fig. \ref{fig:2}, a high concentration of Berry flux exists at the zone boundary with opposite signs between the bottom and the middle bands, but not for the upper band. A competition takes place in the lowest-energy band between thermal occupation factors given by $c_0$ through $c_2$, which favor the occupation of the band minimum at $\vec{k} =0$, and the Berry flux factor concentrated at the zone boundary points with higher energies. For the middle band the two factors cooperate, as the zone boundaries also have the lowest energy.  

An open geometry was used in Ref.~\onlinecite{TMI} to study the edge magnon transport and to deduce the thermal current from the non-equilibrium Green's function formalism. A possible sign change of thermal Hall conductivity with temperature or variations in the parameters of the kagome lattice model Hamiltonian was pointed out in Ref.~\onlinecite{mook}. Both papers~\cite{TMI,mook} were motivated by the magnon thermal Hall experiment on the three-dimensional Lu$_2$V$_2$O$_7$ pyrochlore lattice~\cite{tokura10,tokura12} and obtained the HDMZ Hamiltonian as a two-dimensional projection of the full three-dimensional spin model on the pyrochlore lattice~\cite{tokura10,tokura12}. Following the discovery of thermal Hall effect in the two-dimensional kagome ferromagnetic insulator~\cite{ylee15a}, the authors of Ref.~\onlinecite{lee15} have also carefully looked into the magnetic field dependence of the thermal Hall conductivity in the kagome HDMZ model. 

A drawback of the Holstein-Primakoff approach is that it works poorly in the weakly ordered regime and fails completely for the paramagnetic phase. A paramagnetic state with restored time-reversal symmetry should have zero thermal Hall conductivity anyway, but  by applying magnetic field one can restore thermal Hall transport. There is an uncertainty as to whether the Holstein-Primakoff or the Schwinger boson formalism would capture the physics of Hall effect better for temperatures above the Curie temperature, where thermal Hall effects can be induced by the magnetic field. In the long run, we will benefit from developing a calculation tool more powerful than either of the existing mean-field theories in addressing the high-temperature thermal Hall transport physics of spins.

\section{Experimental Aspects}

\subsection{Two-dimensional topological magnon insulator}

\begin{figure}
\begin{center}
\includegraphics[width=0.4\textwidth]{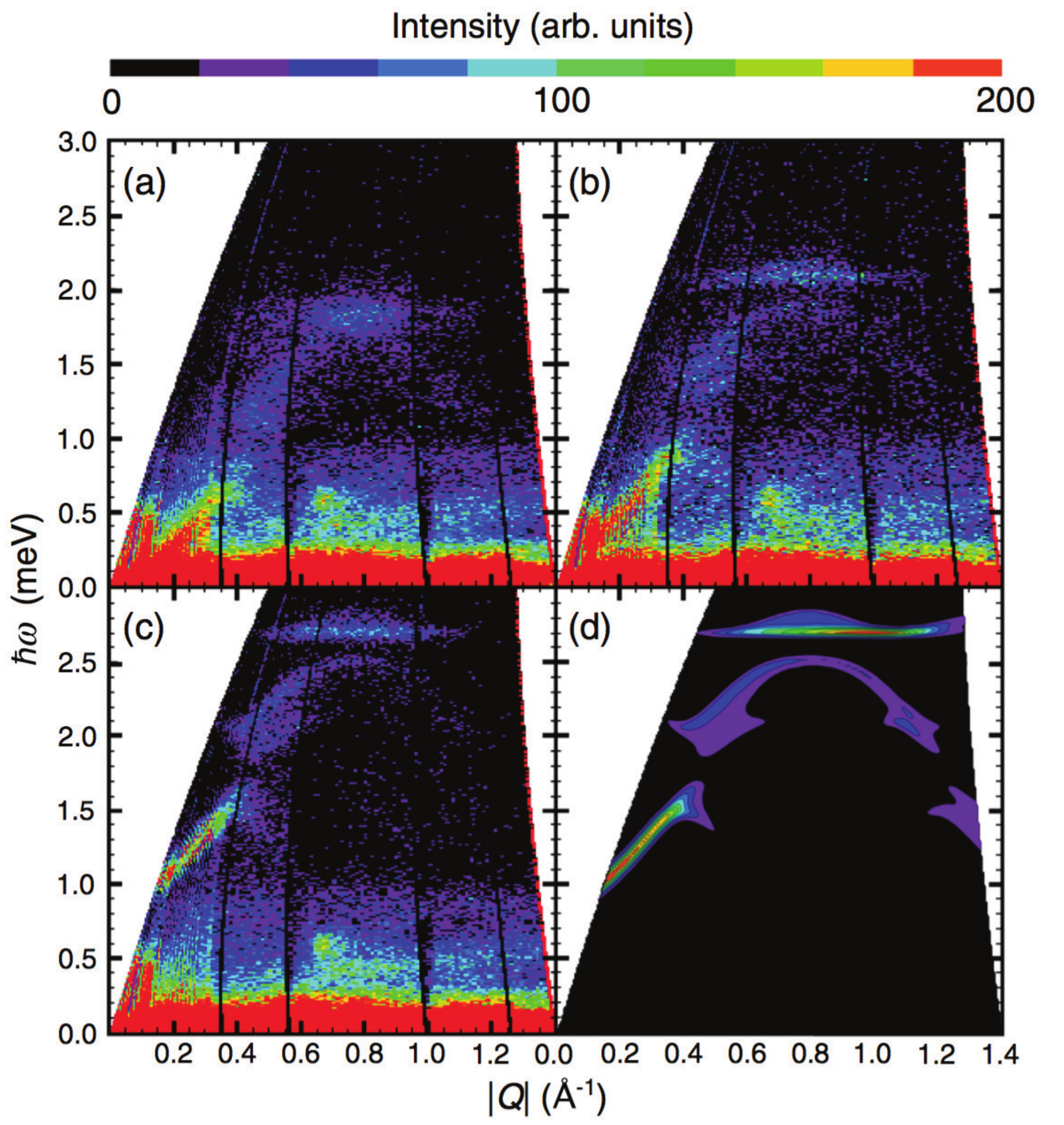}
\caption{Inelastic neutron scattering data showing several magnon bands in Cu(1,3-bdc) under (a) zero, (b) 2T, (c) 7T magnetic fields. Calculated spectra are shown in (d) (reproduced with permission from Ref. \onlinecite{ylee15a}.)}
\label{fig:3}
\end{center}
\end{figure}

\begin{figure}
\begin{center}
\includegraphics[width=0.5\textwidth]{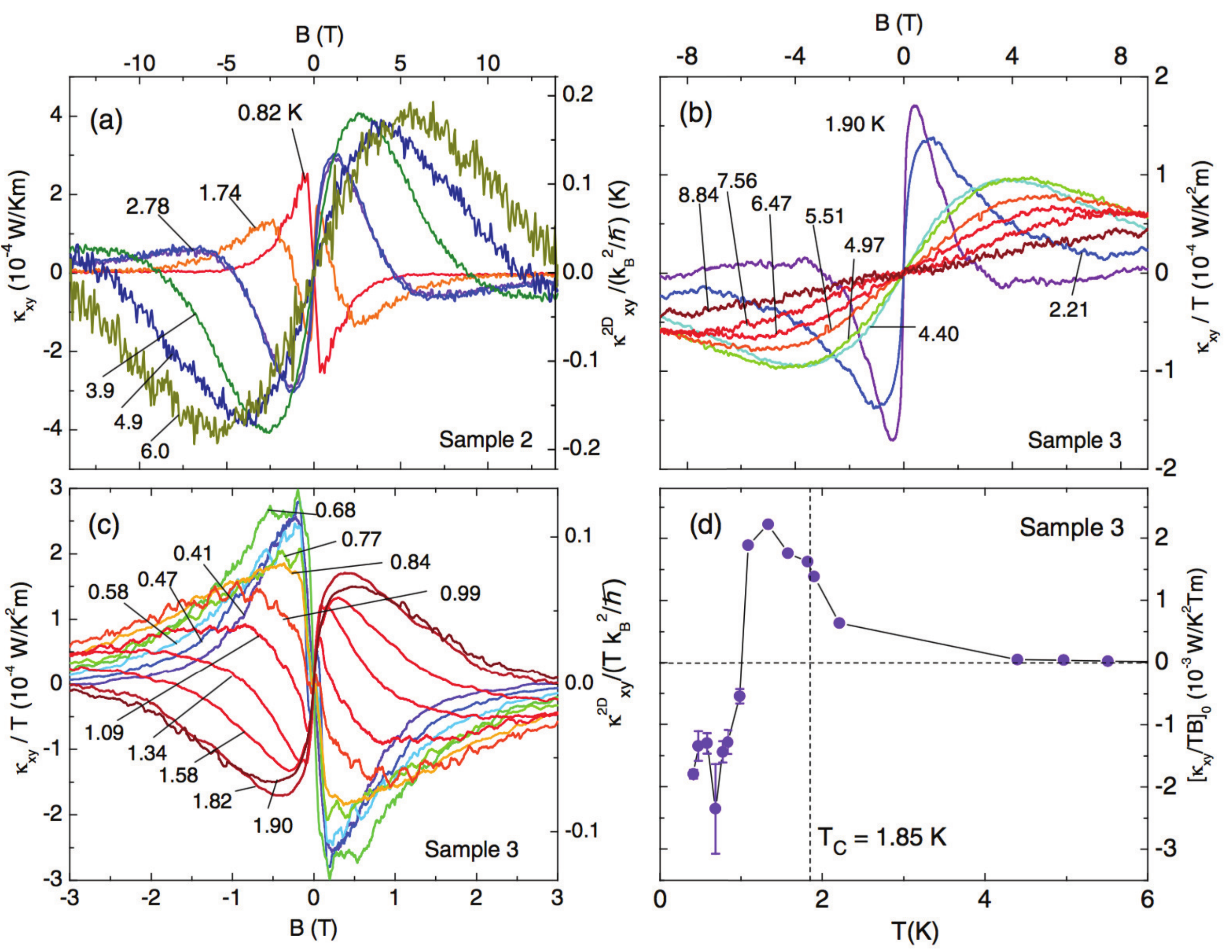}
\caption{Thermal Hall conductivity measurement data on Cu(1,3-bdc).  (reproduced with permission from Ref. \onlinecite{ylee15b}.)}
\label{fig:4}
\end{center}
\end{figure}

The mean-field analysis of the previous section showed that HDMZ ferromagnetic spin model on a kagome lattice is a good example of topological magnon bands, with Chern numbers -1, 0, +1 for successive magnon energy levels. The conditions for realizing the HDMZ model, and thus the topological magnon bands, have been met in the material Cu(1,3-bdc) [bdc=benzenedicarboxylate] synthesized at MIT. Spin-1/2 Cu ions form layers of kagome planes, separated from each other by long, organic (1,3)-bdc molecules. A weak antiferromagnetic inter-layer coupling that exist between the layers can be ignored, allowing us to focus on the dynamics of the two-dimensional kagome lattice. The parameters of the HDMZ Hamiltonian have been deduced by model-fitting the inelastic neutron scattering data such as shown in Fig. \ref{fig:3}, with values $J\approx 0.6$ meV and $D\approx 0.09$ meV $(D/J=0.15)$. The ferromagnetic transition occurs at $T_c=1.77$ K. Due to the small exchange energy scale, applying a few Tesla magnetic field or raising the temperature to a few degrees Kelvin can completely polarize the spins or disorder them, resulting in unprecedented potential to probe magnon dynamics over various field-temperature regimes. Two major outcomes of the experimental investigation by the MIT-Princeton groups are (1) the observation of higher magnon bands from neutron scattering [Fig. \ref{fig:3}], and (2) the observation of thermal Hall effect over wide magnetic field and temperature regimes [Fig. \ref{fig:4}].

The fact that magnons can remain coherent enough in this sample to have the higher-energy bands visible to the neutron probe also point to the potential to explore the Berry flux effects from these same bands through Hall measurements. The thermal Hall conductivity reported in Ref. \onlinecite{ylee15b} and reproduced in Fig. \ref{fig:4} shows both field- and temperature-tuned changes in its sign, reminiscent of competing hole and electron band contributions in the electronic Hall transport but more dramatic than what can be observed in electronic systems. An examination of Fig. \ref{fig:4} tell us that $\kappa_{xy}$ at low fields ($B>0$) is negative ($n$-type) when the measurement temperature is below $T_c$. For $T\gtrsim T_c$, however, low-field $\kappa_{xy}$ becomes positive ($p$-type). Furthermore, the initially $p$-type thermal Hall response obtained at high temperature and low temperature switches to $n$-type when sufficiently strong magnetic field is applied. 

\begin{figure}
\begin{center}
\includegraphics[width=0.5\textwidth]{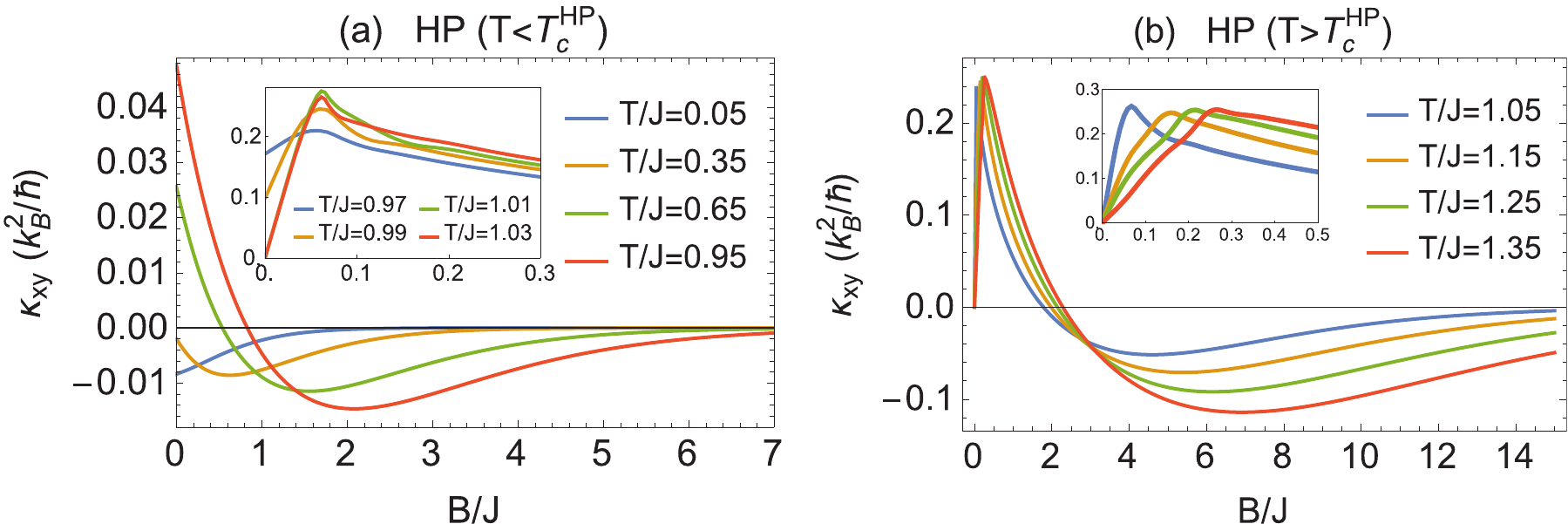}
\caption{Theoretical calculation of the thermal Hall conductivity from the Holstein-Primakoff self-consistent mean-field theory (reproduced with permission from Ref. \onlinecite{lee15}.)}
\label{fig:5}
\end{center}
\end{figure}	

Much of the observed features of $\kappa_{xy}$ in Cu(1,3-bdc) can be understood from a picture of the magnons occupying multiple bands. First, negative values of $\kappa_{xy}$ at the lowest measured temperatures are consistent with the idea of magnons occupying the lowest-energy magnon band where the Chern number is -1. Reduction of the $|\kappa_{xy}|$ value with increasing field is due to the de-population of magnons as the magnon gap increases. A switch to $p$-type Hall transport at low fields and high temperatures implies that higher-energy magnon bands have a substantial occupation as well. The fact that energy levels up to $k_B T$ are sampled with almost equal weights by the $c_2$ function, as seen in Fig. \ref{fig:1}, helps in amplifying the middle and upper band contributions to the thermal Hall conductivity.  The high-temperature $\kappa_{xy}$ turns $n$-type again when a sufficiently large magnetic field de-populate all but the
lowest-energy magnon band. Numerical values of the thermal Hall conductivity from the self-consistent mean-field calculation of the HDMZ model gives a reasonable fit to the measured $\kappa_{xy}$ curves in the case of Holstein-Primakoff boson theory [Fig. \ref{fig:5}]. Schwinger boson mean-field theory generally does a poorer job in fitting the data, even for temperatures above $T_c$. The reason for this is unclear, but one may guess that the applied magnetic field polarizing the spins in the paramagnetic temperature regime validates the Holstein-Primakoff approach once again. 

The convincing evidence from thermal Hall and neutron scattering measurements that magnons in Cu(1,3-bdc) are forming the topological bands also suggests that there will be protected chiral edge modes of magnons in this material. Such topologically protected edge magnon mode might have applications as low-loss, spin-based information carriers.

\subsection{Three-dimensional magnets}

\subsubsection{Pyrochlore ferromagnetic insulator}

Magnon thermal Hall effect was first observed in ferromagnetic pyrochlore insulators
Lu$_2$V$_2$O$_7$ (T$_c \sim $ 70K), Ho$_2$V$_2$O$_7$ (T$_c \sim $
70K), and In$_2$Mn$_2$O$_7$ (T$_c \sim $ 130K). $S=1/2$ V$^{4+}$ ions and $S=3/2$ Mn$^{4+}$ ions occupy the vertices of the tetrahedral network known as the pyrochlore lattice. Magnetic moments order ferromagnetically in the ground state. Overall low-energy spin dynamics is captured by the HDMZ Hamiltonian, with the $\vec{D}_{i\rightarrow j}$ vectors across the adjacent
magnetic sites worked out in Refs. \onlinecite{tokura10,tokura12}. There are four inter-penetrating kagome lattice planes in the
pyrochlore lattice, each parallel to the face of a tetrahedron. For each kagome plane the three-dimensional spin Hamiltonian is reduced to the kagome HDMZ spin Hamiltonian discussed in the earlier section~\cite{TMI,mook}. 

\begin{figure}
\begin{center}
\includegraphics[width=0.35\textwidth]{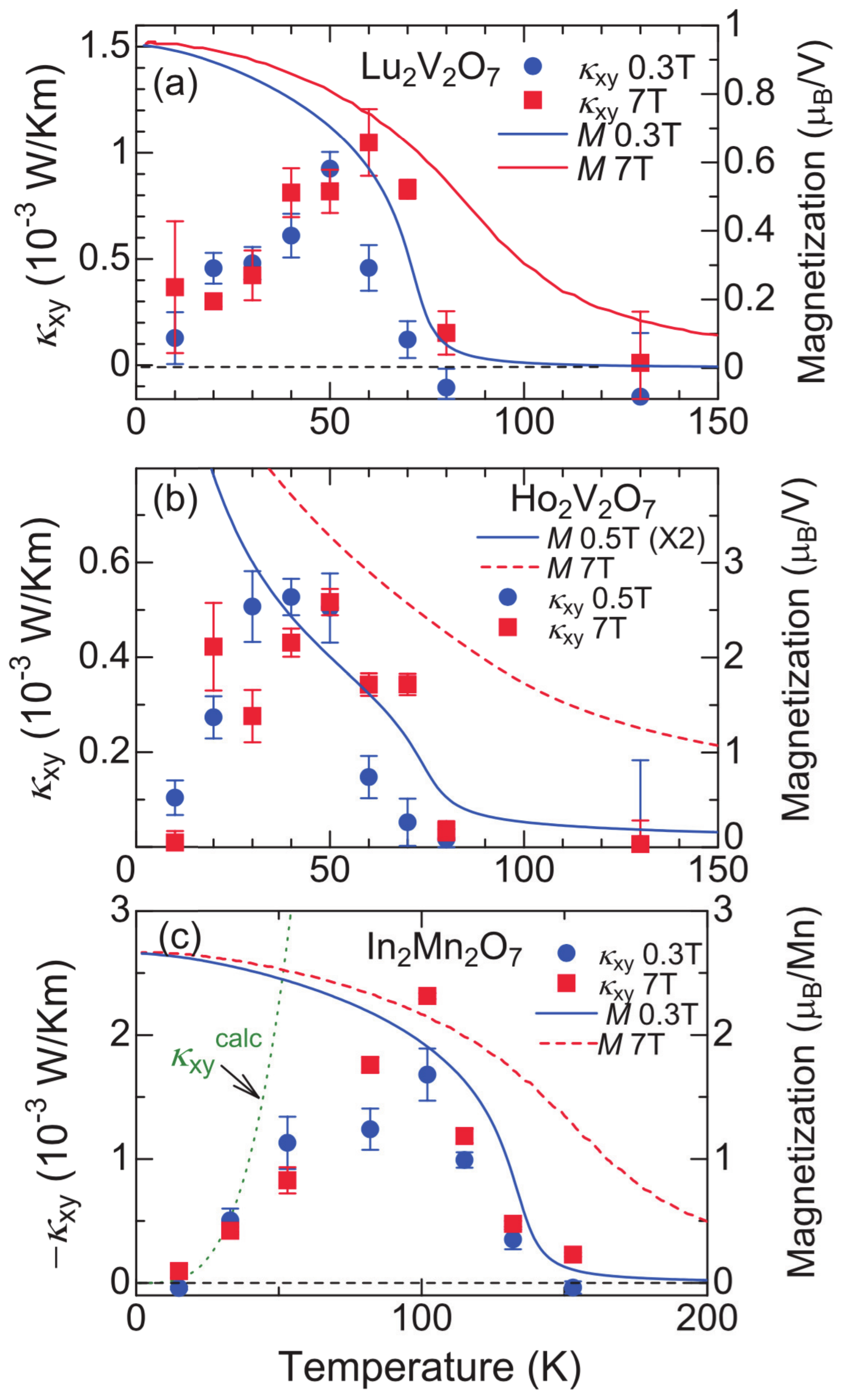}
\caption{Temperature dependence of $\kappa_{xy}$ in several pyrochlore ferromagnets  (reproduced with permission from Ref. \onlinecite{tokura12}.)}
\label{fig:6}
\end{center}
\end{figure}	

The magnetic field dependence of $\kappa_{xy}$ observed in all three compounds bears strong similarity to those of the kagome Cu(1,3-bdc) material at low temperature, and is nor reproduced here. On the other hand,  the field-driven sign reversal of $\kappa_{xy}$ observed in the high-temperature region ($T>T_c$) of Cu(1,3-bdc) is not seen in the pyrochlore material. One explanation for this is that the field strength reaching the exchange energy scale $J$, which is an order of magnitude bigger for the pyrochlore than the kagome compound,
cannot be achieved in the experiment. The other reason might be that there is simple very little Hall response in the paramagnetic region $T>T_c$ for the pyrochlore compound, whereas Hall effects were visible even at several times $T_c$ for Cu(1,3-bdc) [Fig. \ref{fig:4}].
The temperature dependence of $\kappa_{xy}$ in several pyrochlore compounds, at fixed $B$, is reproduced as Fig. \ref{fig:6}. The non-monotonic rise and fall of $\kappa_{xy}$ can be understood from increased thermal occupation of topologically non-trivial, lowest-energy magnon band~\cite{tokura10,tokura12} (rise), followed by the occupation of nearly all bands, resulting in the cancellation of topological effects (fall). Recall that the magnon bandwidth is controlled by $JS^2$ in the self-consistent magnon picture, and $S\rightarrow 0$ as $T_c$ is approached so that the effective magnon bandwidth shrinks considerably. Thermal fluctuation will also reduce the transport. 

Thermal Hall conductivity vanishes completely at $T\sim T_c + 10$K in pyrochlore compounds, while it persists even at temperatures several times $T_c$ for the two-dimensional kagome material~\cite{ylee15a}. One might say offhand that two dimensionality gives rise to more ``correlated" fluctuations. The argument is not entirely convincing, however, given that a sizable magnetic moment is induced by the magnetic field in both materials regardless of the dimensionality, and the dynamics must likely be governed by the magnonic Hamiltonian anyway. A fully self-consistent mean-field calculation for the pyrochlore magnon band, taking into account all of the band contributions to the thermal Hall conductivity, might help us understand the difference in the paramagnetic responses of two- vs. three-dimensional magnetic insulators. 

The perovskite magnetic insulator BiMnO$_3$ having a completely different crystal structure from the pyrochlore was also found to have sizable thermal Hall conductivity~\cite{tokura12}. Construction of a minimal model accounting for the observed thermal Hall signal in BiMnO$_3$ is yet to be taken up.

\subsubsection{Quantum spin ice material}

Thermal Hall transport was recently observed in Tb$_2$Ti$_2$O$_7$, a well-known candidate material to host three-dimensional quantum spin liquid~\cite{ong15}. The trivalent Tb$^{3+}$ ions occupy the sites of the pyrochlore lattice and form, in the crystal field theory jargon, a non-Kramers doublet described as pseudospin $S=1/2$~\cite{gingras14}. Since it is one of the materials believed to be described by the quantum spin ice model, we give a brief overview of it before discussing the results of thermal Hall experiment. A comprehensive review of the theory and experiments on all candidate quantum spin ice materials on the pyrochlore lattice can be found in Ref. \onlinecite{gingras14}. 

A classical spin ice rule is captured by the Ising interaction $H  = J_z \sum_{\langle ij\rangle} S^z_i S^z_j$ for spins at the sites of the tetrahedra, where $S^z_i = +1$ means the spin pointing out from the center of the tetrahedron. Any one of the macroscopic number of configurations where every tetrahedron satisfies the two-in-two-out rule is a classical ground state. One spin flip creates a 3-out-1-in and 1-out-3-in configurations for the adjoining tetrahedra called the magnetic monopole and anti-monopole in the spin ice literature. These monopoles are gapped, however, and do not form a low-energy excitation spectra. Other nearest-neighbor exchange terms as well as the Zeeman term can be included to form the effective spin Hamiltonian~\cite{ross11,onoda11,savary12,sblee12,onoda15}

\begin{eqnarray} &H&=   \sum_{\langle ij\rangle} \Bigl(J_z S^z_i S^z_j - J_\pm (S^+_i S^-_j + S^+_j S^-_i )  \Bigr)   \nonumber \\
&& + J_{z\pm} \sum_{\langle ij \rangle }  S^z_i \left( \xi_{ij} S^+_j + \xi_{ij}^* S^-_j \right) + (i \leftrightarrow j ) \nonumber \\
&& ~~~ - \sum_i (\vec{B} \cdot \hat{e}_i ) S_i^z .\nonumber\\
 \label{eq:qsi} 
\end{eqnarray} 
Details of the structure constants $\xi_{ij}$ can be found in Refs. \onlinecite{ross11,savary12,sblee12}. Unit vectors $\hat{e}_i$ in the Zeeman term serves to project the field $\vec{B}$ to the spin direction $S^z_i = +1$~\cite{onoda15}. 
In the monopole language $J_{z\pm}$ and $J_\pm$ give hopping of the monopole at one tetrahedron to the first and the second neighboring tetrahedra, respectively. A slave-particle solution of the model predicts a U(1) quantum spin liquid forming in the vicinity of the classical spin ice phase, with gapped monopole excitations and a gapless ``photon" excitation~\cite{savary12,sblee12}. Existing efforts at identifying the phase diagram of the effective spin Hamiltonian (\ref{eq:qsi}) have left out the Zeeman term whereas it plays a crucial role in thermal Hall phenomena. A recent exception is Ref. \onlinecite{onoda15} in which classical Monte Carlo calculation based on Eq. (\ref{eq:qsi}) was done to compute specific heat and magnetization. Still, identifying the exact nature of the quantum spin liquid state in the model  (\ref{eq:qsi}) remains a challenging issue. On the experimental side, it is not fully resolved if candidate quantum spin ice materials such as Tb$_2$Ti$_2$O$_7$ and Yb$_2$Ti$_2$O$_7$ fall within the parameter space of the model Hamiltonian (\ref{eq:qsi}). For an extensive discussion on this point, see Ref. \onlinecite{gingras14}. 

\begin{figure}
\begin{center}
\includegraphics[width=0.5\textwidth]{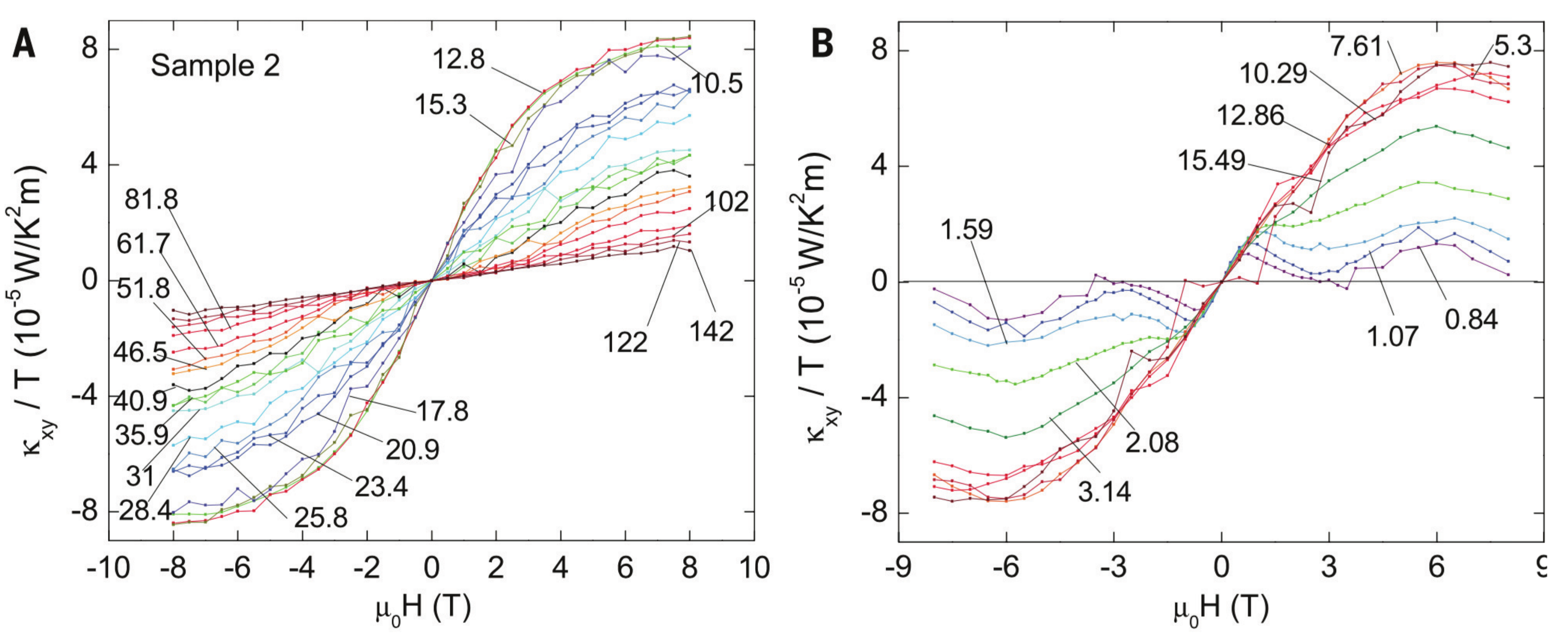}
\caption{Thermal Hall conductivity of Tb$_2$Ti$_2$O$_7$.  (reproduced with permission from Ref. \onlinecite{ong15}.)}
\label{fig:8}
\end{center}
\end{figure}	

With this background in quantum spin ice, let us examine the thermal Hall data taken on the Tb$_2$Ti$_2$O$_7$ compound~\cite{ong15}. A quick scan of $\kappa_{xy}$ reproduced in Fig. \ref{fig:8} shows the field dependence similar to those of the kagome or pyrochlore ferromagnets at high temperature. Sign change in $\kappa_{xy}$ under temperature or field variation is absent in the data, except at the lowest measured temperatures $T\lesssim 1$K where somewhat irregular behavior was observed. As seen in Fig. \ref{fig:8}(b), the temperature-normalized slope $[\kappa_{xy}/TB]_{B\rightarrow 0}$ remains unusually constant for all temperatures below 10K. The other point of significance is the saturation of the temperature-normalized longitudinal thermal conductivity $[\kappa_{xx}/T]_{T\rightarrow 0}$ to a constant value, which remains robust against the applied field variations $|B | \lesssim$ 2T.
Both observations run counter to the standard predictions of the magnon-based theory~\cite{tokura12}. Since the Tb moments do not order down to the lowest measured temperature, the validity of the magnon picture is under scrutiny in the first place. 
A tempting, alternative idea is to view the elementary excitations as fermions that form a Fermi surface and obey the Wiedemann-Franz-type law~\cite{KNL,ong15}. Interestingly, measurement of longitudinal thermal conductivity in another candidate quantum spin ice material Yb$_2$Ti$_2$O$_7$ shows $\kappa_{xx} \sim T^2$ dependence~\cite{matsuda16}, and no observable $\kappa_{xy}$~\cite{matsuda-private}. Understanding why these two putative quantum spin ice materials should show very different thermal responses is a challenging problem. 

Recall that previous observations of thermal Hall effects~\cite{tokura10,tokura12,ylee15a} in magnetic insulators were understood to originate from the Dzyaloshinskii-Moriya interaction in the model Hamiltonian. On the face of it the quantum spin ice Hamiltonian (\ref{eq:qsi}) does not have the DM interaction explicitly built in. As one unravels this Hamiltonian to a global frame, instead of writing the spin operator according to the local quantization axis, all sorts of terms including the anti-symmetric DM exchange will arise~\cite{gingras14}. In a more realistic theory one also needs to take into account the dipolar interaction~\cite{onoda11,onoda15}, which is known to generate the magnon Hall effect under some circumstances~\cite{murakami14}. At this point one does not have a consensus as to the nature of the elementary excitation in the quantum spin ice Tb$_2$Ti$_2$O$_7$ which participates in the thermal Hall transport, nor do we know which mechanism drives it. It might be worthwhile to extend the slave-particle mean-field analysis of Refs. \onlinecite{savary12,sblee12} to include the Zeeman term, and to work out the thermal Hall conductivity formula. 

\section{Chirality-driven Transport in Skyrmionic Matter}

The discovery of Skyrmion phase in a number of chiral ferromagnets spurred a large body of activities in recent years
\cite{pfleiderer,tokura-skyrmion,nagaosa-review,recent-Skyrmion}. The field is still growing at a rapid pace, with the ultimate hope of realizing a Skyrmion-based spintronics device where each Skyrmion can be created, destroyed, transported, detected, and manipulated individually. Aside from device interests, an intrinsic physics tied to the emergent electromagnetism has attracted considerable attention~\cite{zang11}. In short, it means that the spin texture of a Skyrmion acts as a source of magnetic field $\vec{b}$,

\begin{equation} b_\lambda = (1/8\pi) \varepsilon_{\lambda\mu\nu}  \vec{S} \cdot (\partial_\mu \vec{S} \times \partial_\nu \vec{S}) ,\end{equation} 
to the electrons whose spins are tightly locked to the localized moment through the Hund's coupling. Tangible physical effects such as the Hall effect of electrons traveling in the background of Skyrmion lattice have been already observed~\cite{nagaosa-review,recent-Skyrmion}. 

Skyrmion is a quantized version of the spin chirality. Starting from the lattice definition of spin chirality $\chi_{ijk} = \vec{S}_i \cdot (\vec{S}_j \times \vec{S}_k )$, and taking the continuum limit $j = i+a\hat{x}$, $k=i + a\hat{y}$, $a\rightarrow 0$, one finds

\begin{equation} \chi (\vec{r}) = {1\over 4\pi} \vec{S} \cdot (\partial_x \vec {S} \times \partial_y \vec{S} ) \label{eq:chi-continuum} 
\end{equation}
is the spin chirality in the continuum language, whose integral over a two-dimensional closed manifold is an integer. Three-dimensional generalization of the spin chirality is easily obtained by seeing  $\chi (\vec{r})$ as the $z$-component of the chirality vector, and taking cyclic permutations of $(x,y,z)$ indices in Eq. (\ref{eq:chi-continuum}). 

Magnons, like electrons, see the local chirality $\chi(\vec{r})$ as an emergent magnetic field. Each Skyrmion, of the size of tens of lattice constants in practical materials, will be seen as a magnetic flux to the magnon in the same way that a flux tube is seen by the electron in the Aharonov-Bohm scattering. While an electron sees the Skyrmion as carrying one unit of flux quantum, magnon sees it as roughly two units of quanta (more on this point later). Several theory papers have discussed the magnon-Skyrmion interaction in recent years~\cite{jiadong13,loss13,batista14,garst14,iwasaki14,kovalev14,mochizuki14,yt15} in analogy to the Aharonov-Bohm scattering problem.

Derivation of the continuum low-energy theory of the magnon motion in the Skyrmionic spin background, or any kind of static spin background for that matter, follows the same spirit as that of the electronic effective action. The first step is to make a rotation of the spin quantization axis to align with the local moment direction. We discuss how to follow this strategy in the case of the continuum version of the HDMZ Hamiltonian, 

\begin{eqnarray}
  H = \frac{J}{2}\sum_{\mu=1}^d \left( \partial_{\mu} \vec{n} -
  \kappa \hat{e}_{\mu} \times \vec{n} \right)^2 - S  \vec{B} \cdot \vec{n} ,\label{eq:h-bare}
\end{eqnarray}
where $d$ is a spatial dimension, $\vec{n}$ is the local moment normalized to unit length, and $\hat{e}_{\mu}$ is the unit vector in each $\mu$-direction. The DM interaction $D \vec{n} \cdot \vec{\nabla}\times \vec{n}$ follows from expanding the cross terms in the square, and $\kappa=D/J$. For two-dimensional spin systems the index $\mu$ runs up to $d=2$. An orthogonal matrix $R$ rotates $\hat{z}$ to the local equilibrium magnetization direction $\vec{n}_0 = R\hat{z}$\cite{tatara08}. Making the substitution $\vec{n} \rightarrow R\vec{n}$ in Eq. (\ref{eq:h-bare}) yields

\begin{eqnarray}
  H= \frac{1}{2} J \sum_{\mu=1}^d
  \left(\partial_{\mu} \vec{n}  - \vec{a}_{\mu} \times \vec{n}\
  \right)^2 - J S B\, \vec{n}_0 \cdot \vec{n} .
  \label{eq:h-gauged}
\end{eqnarray}
The trade-off one gets from writing the Hamiltonian in the rotated frame is the emergence of gauge fields $\vec{a}_\mu = (a^1_\mu, a^2_\mu, a^3_\mu)$. Each component is related to the underlying ground state spin structure $\vec{n}_0$ through~\cite{liu11,tserkovnyak12,loss13,yt15}

\begin{eqnarray}  & (\vec{\nabla}\times \vec{a}^3)_\alpha = {1\over 2} \varepsilon_{\alpha\mu\nu} \vec{n}_0 \cdot (\partial_\mu \vec{n}_0 \times \partial_\nu \vec{n}_0 ) + \kappa (\vec{\nabla}\times \vec{n}_0 )_\alpha \nonumber \\
& ~~~~~~ + (1-\cos\theta) (\vec{\nabla}\times\vec{\nabla} \phi)_\alpha , \nonumber \\
& -\sum_\mu (a^1_\mu + i a^2_\mu )^2  = e^{2i\phi} \sum_\mu [\hat{\theta} + i \hat{\phi}) \cdot (\partial_\mu \vec{n}_0 - \kappa \hat{e}_\mu \times \vec{n}_0 ) ]^2 . \nonumber\\ \label{eq:a-to-n}
\end{eqnarray}
Various angles in these formulas are coming from the triad of local unit vectors

\begin{eqnarray} 
&& \vec{n}_0 = (\sin \theta \cos\phi , \sin \theta \sin \phi, \cos\theta) , \nonumber \\
&& \hat{\theta}  = (\cos\theta \cos \phi , \cos \theta \sin \phi, - \sin \theta ) , \nonumber \\ 
&& \hat{\phi} = \vec{n}_0 \times \hat{\theta}. \end{eqnarray}

\begin{figure}
\begin{center}
\includegraphics[width=0.35\textwidth]{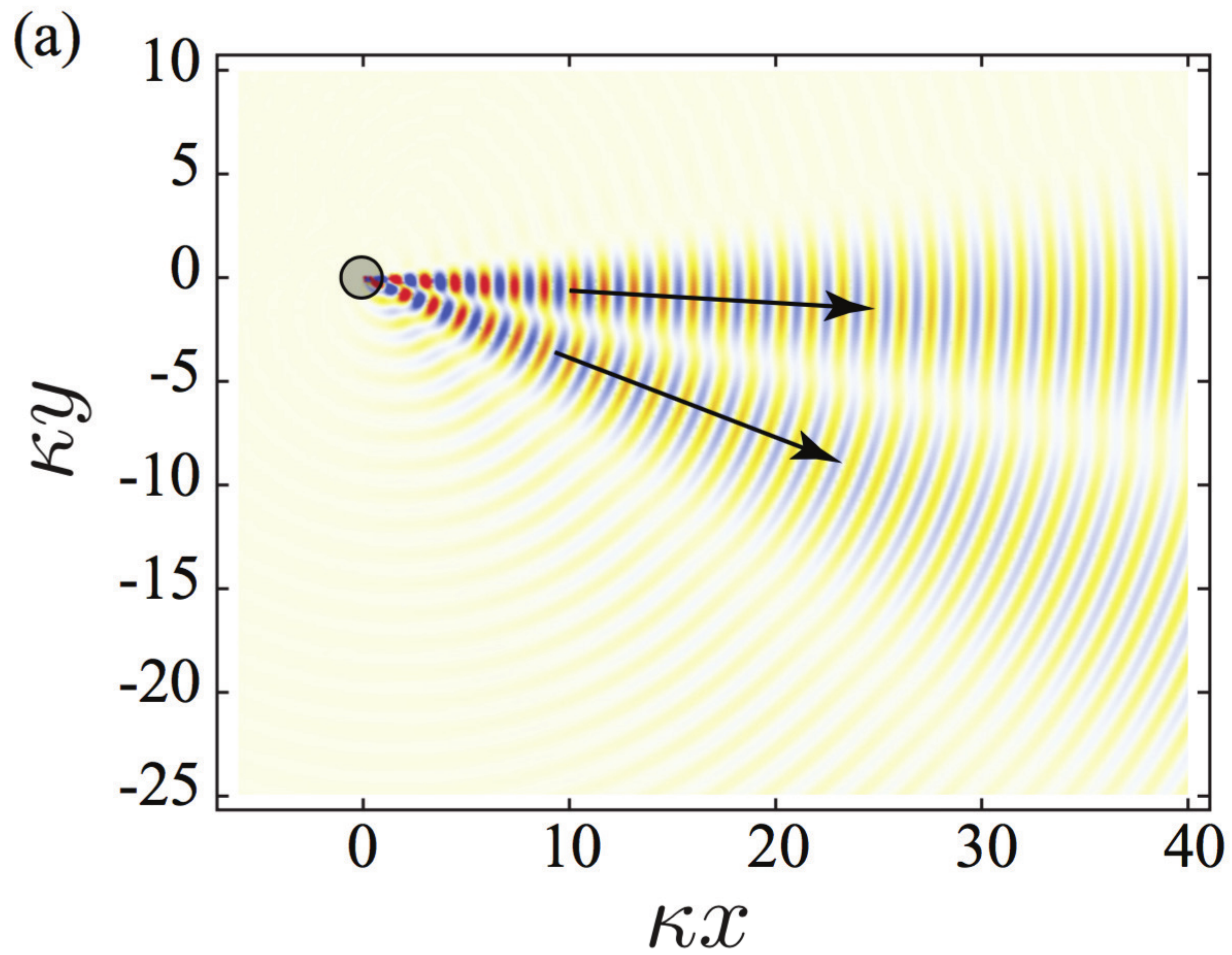}
\caption{Magnon skew scattering off a localized Skyrmion source (reproduced with permission from Ref. \onlinecite{garst14}.)}
\label{fig:9}
\end{center}
\end{figure}

The problem of magnon propagation through the Skyrmion background can be solved in many ways. The most straightforward, but numerically accurate, way is to write down the standard Landau-Lifshitz-Gilbert equation of motion for $\vec{n}$ and integrate it numerically. One can also write down a gauged-LLG equation based on the Hamiltonian (\ref{eq:h-gauged})~\cite{liu11}, and integrate it.  Otherwise, one can derive the magnon Hamiltonian by making the Holstein-Primakoff substitution for $\vec{n}$ in Eq. (\ref{eq:h-gauged})~\cite{yt15}, 

\begin{equation}  {H\over JS} = {1\over 2} B^\dag \begin{pmatrix} - \sum_\mu (\partial_\mu - i a^3_\mu)^2  + m & \Delta^* \\
\Delta & - \sum_\mu (\partial_\mu + i a^3_\mu)^2  + m \end{pmatrix} B , \label{eq:h-magnon} \end{equation}
where $B=\begin{pmatrix} b \\ b^\dag \end{pmatrix}$, and

\begin{eqnarray} 
& m = \vec{B} \cdot \vec{n}_0 - (1/2) \sum_\mu [(a^1_\mu)^2 + (a^2_\mu)^2 ] ,  \nonumber \\
& \Delta = -(1/2)\sum_\mu (a^1_\mu + i a^2_\mu)^2. 
\end{eqnarray}
The magnon Hamiltonian or some analogous form of the continuum model has been used to address the magnon-Skyrmion scattering problem~\cite{garst14,iwasaki14,yt15}. The essential idea for all the theories is that the magnon sees an effective flux in a Skyrmion through the $\vec{a}^3$ term in the minimal coupling. The first term for $\vec{\nabla}\times\vec{a}$ in Eq. (\ref{eq:a-to-n}) indeed suggests that a Skyrmion is seen to carry two units of flux quanta for magnons. A more careful treatment, however, finds two more terms contributing to the curl of $\vec{a}^3$. Figure \ref{fig:9} shows the skewed trajectory of a magnon hitting off a Skyrmion, taken from Ref. \onlinecite{garst14}. Other works~\cite{iwasaki14,yt15} identified a similar scattering process. 

The idea of emergent magnetic field for magnons can be carried over naturally to the Skyrmion lattice. On average magnons would see the Skyrmion lattice as uniform magnetic field and form Landau levels. The inhomogeneous, but periodic part of the emergent field perturbs the Landau level picture and gives rise to a set of dispersive bands carrying nonzero Chern numbers~\cite{loss13}. 
A recent work has derived the magnon spectra in the bulk and the edge modes for the triangular lattice of Skyrmion~\cite{molina}.
The thermal Hall conductivity of magnons in the Skyrmion background can then be calculated using the formula worked out in Eq. (\ref{eq:kappa-MF})~\cite{loss13}. It will be possible, then, to observe magnon thermal Hall effect in the insulating Skyrmion crystal such as Cu$_2$OSeO$_3$~\cite{seki12}. 

The simple picture described above, however, does not take into account the ``back action" of the Skyrmion in response to the magnon. As noted by several authors~\cite{jiadong13,batista14,kovalev14,mochizuki14}, magnon current creates a Skyrmion motion both against the magnon flow (longitudinal back-action) and perpendicular to it (transverse back-action). The longitudinal part of the Skyrmion back-action can be pictured as a horse racing on an extended treadmill. As the horse (magnon) gallops forward, the treadmill (Skyrmion) is pushed backwards in keeping with the momentum conservation. The spin transfer torque is mediating the exchange of momentum between the two parts of the spin system. Measurement of the thermal conductivity in an insulating Skyrmion crystal can be quite hard to interpret, in reality, due to the fact that both magnons and Skyrmions will be set in motion in response to the thermal gradient. In particular the longitudinal part of the thermal transport 
seems to have one positive contribution from magnons flowing from high to low temperature ends, and another {\it negative} contribution from the Skyrmions which move towards the higher temperature end of the contacts! 

A smart way to avoid the complications of the longitudinal back-action of the Skyrmion, and to focus sharply on the magnon Hall transport in the Skyrmion crystal, is to confine the Skyrmions in a restricted geometry.  A laser beam, for instance, can create a temperature gradient that points radially from the center of the laser spot where it would be hottest. The magnon current generated by the temperature gradient has a Hall component because it sees the Skyrmion lattice background as magnetic field, directed tangentially to the droplet. The tangential magnon flow will then ``kick back" the Skyrmions and set them in a Ferris wheel motion, and a Feynman's ratchet will be realized. The ratchet motion of the Skyrmion crystal beautifully observed in the Lorentz TEM experiment (where the laser spot is replaced by an electron beam) is interpreted in this way~\cite{mochizuki14}, and amounts to the observation of magnon Hall effect in the Skyrmion crystal. It is amusing to speculate that the same novel phenomena we are discussing in the context of magnon-Skyrmion scattering may arise in the phonon-vortex scattering in superfluids as well. The Lagrangian for the vortex motion shares the same Berry phase term as the Skyrmion~\cite{stone}. 

The ability to drive Skyrmions by spin waves (magnons) is of importance from the practical importance as well. Following the insightful suggestion in Ref. \onlinecite{fert}, the authors of Ref. \onlinecite{xichao} considered the magnon-driven motion of Skyrmion on a nano-sized strip whose width is a few times the Skyrmion radius. Interestingly, the Skyrmion was driven forward by the magnon and not backwards as in Refs. \onlinecite{jiadong13,batista14}. Somehow the confining potential of the nanostrip plays a role in the Skyrmion motion as well. 

\section{Summary and Outlook}

Hall-like transport phenomena of spin excitations in insulating magnets have been reviewed with emphasis on ``paramagnetic" thermal Hall effects seen in two-dimensional ferromagnetic insulator Cu(2,1-bdc) and putative quantum spin ice Ti$_2$Ti$_2$O$_7$. Both cases showed a substantial thermal Hall conductivity over temperatures significantly above the ordering temperature $T_c$ (if ordering exists at all, in the case of Ti$_2$Ti$_2$O$_7$). Formal theories of spin thermal transport without assuming magnetic order, hence different from the magnon picture, have been outlined. An attractive venue to look for spin thermal transport in the absence of magnetic order is the spin liquid insulator. Thermal Hall conductivity measurement on the triangular spin liquid material EtMe$_3$Sb[Pd(dmit)$_2$]$_2$ did not show a discernible thermal Hall signal~\cite{matsuda10}. On the other hand, another candidate spin liquid material Cu$_3$V$_2$O$_7$(OH)$_2 \cdot$2H$_2$O (Volborthite) was recently found to have finite $\kappa_{xy}$ over the temperature region where significant spin-spin correlations exist~\cite{matsuda-PNAS} but no magnetic order is present. 

Another promising candidate spin system to look for thermal Hall effect is the antiferromagnetic insulator on a frustrated lattice (e.g. kagome) with DM exchange. The Heisenberg exchange alone leads to planar spin order without spin chirality, but the addition of suitably chosen DM vector is known to cant the spins out of plane, leading to an umbrella-like structure that has finite spin chirality~\cite{lacroix}. It is possible, according to known rules of classical statistical mechanics that Ising order is harder to destroy than a continuous one, to have a phase where the long-range antiferromagnetic order is destroyed but the Ising-like spin chirality order survives. Such will be a phase in which time-reversal symmetry remains spontaneously broken, where one can expect the thermal Hall effect without applying the magnetic field. A recent report~\cite{broholm16} of the spin-canted order in the antiferromagnetic insulator in Nd$_3$Sb$_3$Mg$_2$O$_{14}$, in which Nd $J=1/2$ moments sitting on kagome sites were found to be in a canted antiferromagnetic phase, is a promising material avenue to look for ``spontaneous" thermal Hall effect in the paramagnetic regime. A large canting angle and the estimate of a large DM interaction $D/J\approx 1$, together with a significant amount of spin-spin correlations observed above the Neel temperature, further suggests that a thermally driven spin liquid phase may well exist in this material. Observation of the thermal Hall signal in such a phase would further highlight the spin-chirality origin of thermal transport in magnetic insulators. 

On the theoretical front, calculation of the fully spin-based thermal Hall conductivity formula (\ref{eq:full-kappa-E}), without the recourse to mean-field theory, will be challenging but interesting to carry out. Even the classical simulation of spin dynamics can give the thermal Hall coefficient through the well-known Green-Kubo formula~\cite{green,kubo57}

\begin{equation}  \kappa_{ab}^E = {1\over V k_B T^2 } \int_0^\infty \langle I^E_a (t) I^E_b (0) \rangle_{ens.} ~ dt , \end{equation}
where the current operators are now written in terms of classical spin variables, will be interesting to carry out~\cite{zotos}.

We also reviewed the chirality-driven magnon transport phenomena in the Syrmionic matter. Emergent gauge field induced by the non-trivial spin background results in the local chirality functioning as the magnetic field. Consequently, the itinerant magnon feels the emergent magnetic field, resulting in such phenomena as the skew scattering off a localized Skyrmion and Hall-like thermal transport. A three-dimensional analogue of the Skyrmion-driven Hall effect can take place if the magnon band structure assumes the Weyl form, with a linear dispersion around points in the three-dimensional Brillouin zone and a non-zero monopole number. An interesting proposal for Weyl magnons and the thermal Hall effect in such system was proposed recently~\cite{gang-chen}.

%\begin{table}
%\caption{List of options for paper types.}
%\label{t1}
%\begin{center}
%\begin{tabular}{ll}
%\hline
%\multicolumn{1}{c}{Option} & \multicolumn{1}{c}{Paper type} \\
%\hline
%\verb|ip| & Invited Review Papers \\
%\verb|st| & Special Topics \\
%\verb|letter| & Letters \\
%\verb|fp| & Full Papers \\
%\verb|shortnote| & Short Notes \\
%\verb|comment| & Comments \\
%\verb|addenda| & Addenda \\
%\verb|errata| & Errata \\
%\hline
%\end{tabular}
%\end{center}
%\end{table}

%\begin{acknowledgment}

\acknowledgments We are indebted to Robin Chisnell, Hosho Katsura, Young S. Lee, and P. Ong for valuable discussions on various aspects of thermal Hall effect. Patrick A. Lee collaborated with the authors on the spin theory of thermal transport and shared numerous insights on the theme. JHH thanks Shuichi Murakami for discussion on several cross-related themes and for the invitation to write this review. HYL is supported by the NRF grant (No.2015R1D1A1A01059296).

%\end{acknowledgment}

%\begin{verbatim}
%\profile{Jung Hoon Han}{was born in Seoul, South Korea, in 1969. After graduating from the physics department of Seoul National University he joined the graduate school at the University of Washington as a student of professor David Thouless.  After two stints as post-doctoral fellows at Asia Pacific Center for Theoretical Physics and at UC Berkeley, he joined the faculty of Konkuk university in 2001 and later Sungkyunkwan University, where he stayed from 2003 to the present.}
%\profile{Hyunyong Lee}{was born in Seoul, South Korea, in 1984. He earned his Batchelor's degree from the physics department of Sungkyunkwan University and earned his PhD at Pohang University of Science and Technology under the supervision of Stefan Kettemann. After finishing his doctorate, he came back to Sungkyunkwan University as a post-doctoral fellow in Jung Hoon Han's group since 2014. Apart from interests in spin transport physics, he has recently been doing work on tensor network theory.}
%\end{verbatim}

\end{document}